# First and second order all-optical integrating functions in a photonic integrated circuit

**Marcello Ferrera,**[1,2,*] **Yongwoo Park,**[1] **Luca Razzari,**[3] **Brent E. Little,**[4] **Sai T. Chu,**[5] **Roberto Morandotti,**[1] **David J. Moss,**[6] **and José Azaña**[1]

[1]*INRS - Énergie, Matériaux et Télécommunications, 1650 Blvd Lionel Boulet, Varennes (Québec), J3X1S2, Canada*
[2]*University of St Andrews, School of Physics and Astronomy, North Haugh, St Andrews, Fife, Scotland, UK, KY169SS,*
[3]*IIT, Istituto Italiano di Tecnologia, Via Morego, Genova, 301663, Italy*
[4]*Infinera Ltd, 169 Java Drive, Sunnyvale, California 94089, USA*
[5]*City University of Hong Kong, Department of Physics and Material Science, Tat Chee Avenue, Hong Kong, China*
[6]*CUDOS, School of Physics, University of Sydney, Sydney, NSW 2006, Australia*
*\*mf39@st-andrews.ac.uk*

**Abstract:** We demonstrate all-optical temporal integration of arbitrary optical waveforms with temporal features as short as ~1.9ps. By using a four-port micro-ring resonator based on CMOS compatible doped glass technology we perform the 1$^{st}$- and 2$^{nd}$-order cumulative time integral of optical signals over a bandwidth that exceeds 400GHz. This device has applications for a wide range of ultra-fast data processing and pulse shaping functions as well as in the field of optical computing for the real-time analysis of differential equations.

## 1. Introduction

All-optical integrated circuits bring the promise of overcoming the limitations of electronic circuits in terms of processing speed, power consumption, and isolation for possible sources of data corruption [1]. One of the key factors that has made electronics a dominant technology for the past 60 years is the ability to perform quite a broad range of fundamental operations using only a small number of basic building blocks [2-3]. These include temporal differentiators and integrators, which have played a major role in the development of the first electronic ALUs (Arithmetic Logic Units). Because of its superior signal to noise ratio, the integrator is often preferred to the differentiator [4-5] for high frequency applications.

A temporal integrator is a device that is capable of performing the time integral of an arbitrary input waveform [6], characterized by its ability to store energy in one form or another (eg., optical, electrical). In electronics, a simple parallel plate capacitor can achieve this due to its inherent capability to store electrical charge, whereas to accomplish this in the optical domain, photons need to be stored and localized in the same fashion as a capacitor accumulates electrical charge – a well known and difficult challenge for photonics.

The realization of an optical integrator in integrated, or monolithic, form would represent a fundamental step for many ultra-fast data processing applications, including photonic bit counting [7], pulse waveform shaping [1,7-8], data storage [9, 10], analog-to-digital conversion [11], and real time computation of linear differential equations [12]. For many of these applications, and particularly the latter, the ability to perform optical integration even just up to second order would be extremely useful. Indeed, a very broad family of phenomena in applied physics, engineering, and biology can be modeled by second order differential equations [13]. In addition, the intrinsic sensitivity of optical integrators to the phase of the signal is in stark contrast to the electronic counterpart that operates on real valued temporal signals. This phase sensitivity can in principle allow for the realization of a new class of functions such as complex optical pulse shaping methods and optical memories [7, 9].

Recently, we demonstrated ultra-fast all optical integration in a monolithic passive device, based on a high-quality factor ($Q>10^6$) micro-ring resonator [14] in a CMOS compatible platform. We achieved 1$^{st}$-order temporal integration of complex field optical waveforms, with a time resolution of ~8ps over an integration window exceeding 800ps, for an overall time-bandwidth product (TBP>100) well beyond the reach of electronics. However, a limitation of that device was its reduced throughput (0.015%) imposed by its very narrow resonance linewidth.

In this work, we explore the trade-off between integration bandwidth and overall energy efficiency by performing all-optical integration in a micro ring resonator with a reduced Q factor (~65.000). By lowering the Q factor, we gain multiple benefits such as increasing the time resolution as well as the efficiency of the device, together with reducing its physical dimensions. Of course, this comes at the expense of a shorter integration time window since a smaller Q results in faster energy dissipation for each round-trip. In addition, we investigate the possibility of implementing 2$^{nd}$-order integration using the same device by directing the output to the secondary input of our four-port device. By doing so, we achieve 1$^{st}$- and 2$^{nd}$-order temporal integration of arbitrary input waveforms with time features down to ~1.9ps, with an input to output power efficiency of 1.5%, an integration bandwidth exceeding 400GHz, and an integration time window of ~12.5ps. This device offers significant promise for integrated ultrafast optical information processing, measurement and computing systems.

## 2. Theory

The approach to realizing an optical integrator relies on emulating the spectral transfer function of an ideal 1$^{st}$-order integrator (linear optical filter), that is proportional to the following:

$$H(\omega) = \frac{1}{j(\omega - \omega_0)} \quad (1)$$

where $\omega_0$ is the central frequency of the optical signal to be processed and $\omega$ is the angular frequency variable. This characteristic stems from the fact that in the time domain, the temporal impulse response of the ideal 1$^{st}$-order integrator $h_{int}(t)$ must be proportional to the unit step function u(t), which is defined by:

$$u(t) = \begin{cases} 0 \text{ for } t < 0; \\ 1 \text{ for } t \geq 0; \end{cases} \quad (2)$$

It is immediately clear that in order to have such a response, an optical device has to be capable of storing energy (photons). We now compare $H(\omega)$ to the spectral transfer function of an optical resonant cavity, such as a Fabry-Perot resonator, in which $\omega_0$ is one of its resonant frequencies. From Fig. 1, we note that these two curves almost overlap for a frequency range that significantly exceeds the cavity linewidth (the full width at half maximum (FWHM) of the resonance). Thus, in photonics, the behavior of a 1$^{st}$-order optical integrator can be emulated by means of an optical resonant cavity.

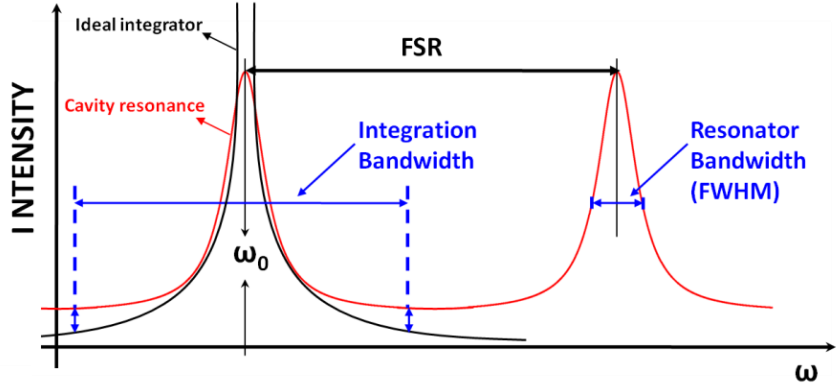

Fig. 1. Integrator transfer function, showing a comparison between the spectral transfer function of an ideal integrator (black curve) with that of a Fabry-Perot cavity (red curve) in which one resonance matches the integrator operative frequency $\omega_0$. The figure shows the main discrepancies between these two curves around the operative frequency and at the lobes of the characteristic resonance. It also shows (blue lines) how largely the operation (processing) bandwidth of the resonator can exceed its linewidth when it is used as integrator.

However, this technique has some limitations. From Fig. 1 we see that the two transfer functions differ from each other (i) at the lobes of the Lorentzian function that approximates the resonance spectral shape (over the cavity free spectral range) and (ii) around $\omega_0$ where the ideal integrator diverges to infinity. To understand this we consider the temporal impulse response of a resonant cavity (e.g. Fabry-Perot) for an input pulse centered at $\omega_0$ and with a time duration longer than the cavity round-trip propagation time:

$$h_{cavity}(t) = \exp(-K \cdot t) \cdot u(t) \qquad (3)$$

where $K=(1/T)\ln(r^2\gamma)$, r is the field reflectivity (e.g. in the Fabry-Perot mirrors), T the round-trip propagation time in the cavity, and $\gamma$ the gain in the cavity medium ($\gamma>1$ in active media and $<1$ in lossy materials), respectively. Notice that the field reflectivity in a Fabry-Perot cavity is equivalent to the cross-coupling field coefficient in a ring resonator. Eq. 3 implies that in a resonator, for each round-trip, part of the propagating light is coupled out, causing an exponential decay of the stored light over time. To ensure that a cavity behaves as a temporal integrator, the condition $r^2\gamma=1$ (i.e. $K = 0$) must be met, which can be achieved in one of two ways: i) compensating both propagation and coupling losses by using an active device with gain ($\gamma>1$; $r<1$; $r^2\gamma\approx1$) or ii) using an ultra low-loss material platform ($\gamma\approx1$) which is also mature enough to allow for a very high mirror reflectivity ($r\approx1$). However, the first strategy implies the use of gain in the cavity and this introduces other limitations such as limited processing speed <20GHz as well as increased signal to noise ratio, in addition to adding fabrication steps that may not be CMOS compatible [12, 15]. On the other hand, typical passive photonic integrators such as those based on fiber Bragg gratings (FBG), either suffer from a very limited integration time window [16, 17] or require a reflectivity approaching 100% [18], which is quite a stringent requirement. In this work we show that, despite some of the limitations inherent in using a resonant optical cavity to approximate an ideal optical integrator (e.g. its limited integration time window), we can achieve a level of performance that in many respects (e.g. integration speed, I/O power efficiency, etc) succeeds much better than other approaches (FBGs, etc).

Generalizing the arguments of the previous paragraph, an $N^{th}$-order (with $N = 1, 2, 3 \ldots$) optical temporal integrator is a device capable of calculating the $N^{th}$ time integral of an arbitrary optical temporal waveform. Since these devices are linear, in general an $N^{th}$-order integrator can be implemented by concatenating N identical $1^{st}$-order integrators. Here, we demonstrate both $1^{st}$-order and $2^{nd}$-order optical integration, achieved using a single micro-ring resonator. We obtain $2^{nd}$-order integration by passing the optical signal through the same resonant cavity twice, by properly re-circulating the signal through the device, as detailed in the following section.

## 3. Experiment

The device is a doped silica glass high index-contrast micro ring resonator with a Q~65.000 and a FSR~575GHz. The waveguide cross section is 1.5x1.45μm$^2$ while the ring radius is 47.5μm. The ring geometry, together with its operating principle as well as an SEM image of the device cross section, are shown in Fig. 2.

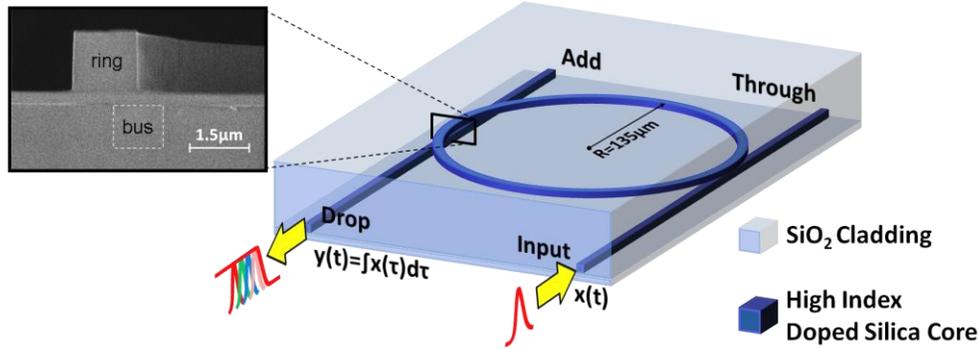

Fig. 2. Schematic of the all-optical integrator. The image on the left shows a SEM image of the micro-ring resonator cross section (taken prior the deposition of the SiO$_2$ cladding). The figure on the right depicts the ring resonator device and illustrates the overall I/O scheme of the cavity when it operates as integrator.

The glass films were deposited by plasma enhanced chemical vapor deposition (PECVD) while the waveguide pattern was implemented using photolithography and reactive ion etching. The fabrication process is CMOS compatible with no need for post annealing at high temperature to reduce losses. The most important characteristic of this device that enabled us to exploit our cavities as all-optical integrators is their extremely low propagation loss, at less than 0.06dB/cm [19, 20].

The complete experimental set-up is shown in Fig. 3 for both the 1$^{st}$- and the 2$^{nd}$-order integration. In both cases we used a mode-locked fiber laser (Pritel) as the signal source, that emitted nearly Gaussian pulses with a central wavelength in resonance with the ring @ 1553.37nm. The optical pulses had a time duration of ~1.9ps and repetition rate of 16.9MHz. The laser output was split into two by means of a 50/50 fiber splitter, one arm of the splitter directed the beam towards a Michelson interferometer that shaped the input waveforms to be processed. The second arm of the splitter was a reference line used to obtain a phase sensitive spectrum of the device output in order to reconstruct the temporal output [21]. In the reference arm the signal was first delayed and then filtered through a polarizer - the polarization was set to TM to align the signal with the selected TM resonance @ 1553.37nm. The reference signal was then interfered with the device output by means of a 50/50 fiber coupler and the result recorded on an optical spectrum analyzer (OSA). The resulting spectra were then used to retrieve the complex-field (amplitude and phase) temporal information of the cavity (integrator) output by using Fourier transform spectral interferometry (FTSI) [22-23]. The latter being the most commonly used algorithm in spectral interferometry due to its simplicity in terms of experimental set-up and data analysis. With this technique, the required spectral resolution, in order to retrieve the temporal profile of an optical pulse, is approximately $\delta\omega \sim 1/(5\cdot\Delta t)$ where $\Delta t$ represents the pulse time duration. This one corresponds in the wavelength domain to ~0.1nm for a 2ps pulse, which is at least one order of magnitude lower than the resolution available in our set-up. In our previous work [14] the output of the integrator was simply detected with a 50GHz bandwidth detector/sampling oscilloscope. In this work, however, the integration time of the device was much shorter (see below) and in fact was on the order of the detector response time. Therefore we adopted the FTSI based measurement approach as a means of achieving a faster detection response time - much faster than the integration response time. Using this system we were able to directly measure the output pulses of the laser (1.9ps pulsewidth – see below) demonstrating that our measurement response time was much faster than the integrator response time. We estimate that the time resolution of our measurement system is ~500fS.

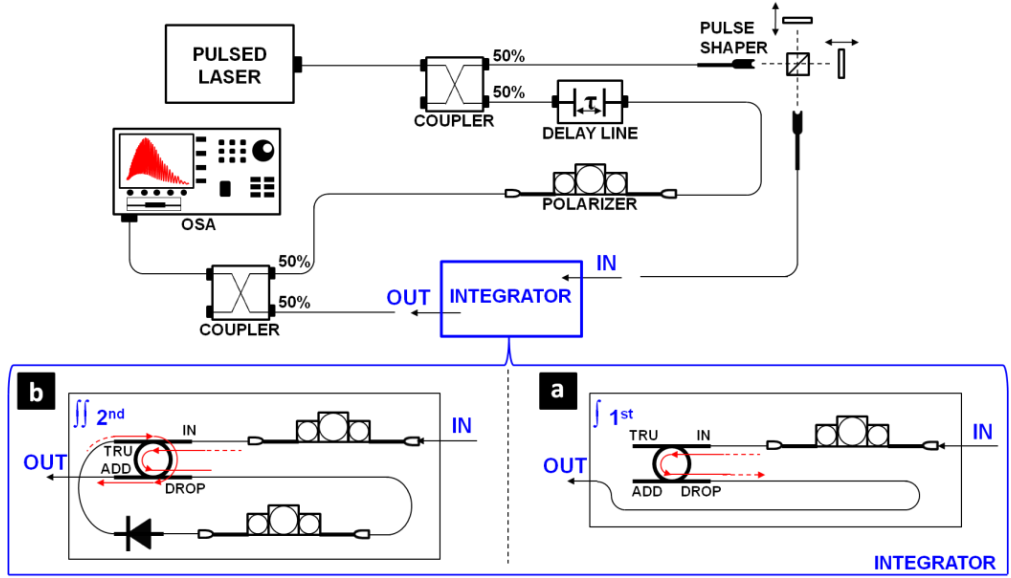

Fig. 3. Experimental set-up. Block (a) and block (b) in the blue box represent the 1st- and the 2nd-order integration modules respectively. During the experiment a mode-locked fiber laser (Pritel) emitted nearly Gaussian pulses with a central wavelength in resonance with the ring @ 1553.37nm. The optical pulses had a time duration of ~1.9ps and repetition rate of 16.9MHz. The laser output was split into two by means of a 50/50 fiber splitter, one arm of the splitter directed the beam towards a Michelson interferometer that shaped the input waveforms to be processed. The second arm of the splitter was a reference line used to obtain a phase sensitive spectrum of the device output in order to reconstruct the temporal output [21]. In the reference arm the signal was first delayed and then filtered through a polarizer - the polarization was set to TM to align the signal with the selected TM resonance @ 1553.37nm. The reference signal was then interfered with the device output by means of a 50/50 fiber coupler and the result recorded on an optical spectrum analyzer (OSA). The resulting spectra were then used to retrieve the complex-field (amplitude and phase) temporal information of the cavity (integrator) output by using Fourier transform spectral interferometry (FTSI) [22-23].

As we can see from Fig. 3, in the case of single integration (block-a), the device simply consisted of the micro ring resonator plus an input polarizer used to filter out the TE polarization from the optical signal before it was launched into the ring input port. For the double integration scheme (block-b), the 1$^{st}$ order integration signal coming out from the drop port was re-directed to the through port to be integrated for a second time. In this configuration, a second polarizer was placed right after the through port to reset the polarization to TM since the propagation inside the ring slightly changed the signal polarization state. For the 2$^{nd}$-order integration scheme (Fig. 3-block-b) an optical isolator was necessary in order to block any residual signal that was not completely coupled into the ring from the input port, since the spectra of the input waveforms exceeded the resonator linewidth.

The optical power spectra of both the input and output of the device are shown in Fig. 4-a, while the corresponding temporal measurements (normalized field amplitude Vs time in ps) are shown in Figure 4-b.

In Fig. 4-a, the I/O characteristics were obtained by recording the power spectra of both the source waveform (input) and the signal at the drop port (output) when the operative ring resonance was excited at the input port.

Fig. 4-b reports both the experimental (black solid curve - retrieved by FTSI) and the theoretical (blue dashed line) device output that correspond to the 1st order integral of the input signal as generated by the laser. The red curve in the inset represents the temporal profile of the optical pulses generated by the laser.

It is worth mentioning that the 1st-order integral reported in Figure 4-b is approximately equal to the temporal impulse response of the system, since the spectral content of the pulse very nearly matches the processing bandwidth of the device. This information was preliminary obtained first by fitting the power spectrum of an ideal integrator with that one of the correspondent (experimentally measured) operative resonance and then estimating the frequency range within which these two curves almost overlap. The resulting integration bandwidth was estimated to be >400GHz a value which is comparable to the FSR of the device (~575GHz) but much wider than the resonator linewidth (~3GHz).

In the same graph, the black solid curve represents the experimentally measured temporal output of the device, while the dashed blue line is the theoretically calculated integral of the input pulse. The input signal to the integrator (the pulse-train generated by the laser) shown in Fig. 4-b was measured by blocking the output of the integrator, thus shielding one arm of the interferometer (Inset Fig. 4-b), and then using the FTSI based method described above. The pulse temporal profile was obtained by fitting the FTSI-obtained waveform with a Gaussian profile and then measuring the full width at half maximum, which resulted in a pulse-width of ~1.9ps (see inset Fig. 4-b). Finally, from Fig. 4-b, we measured the device integration time window to be as long as ~12.5ps, defined as the decay time of the integrator impulse response to reach 90% of the maximum field amplitude.

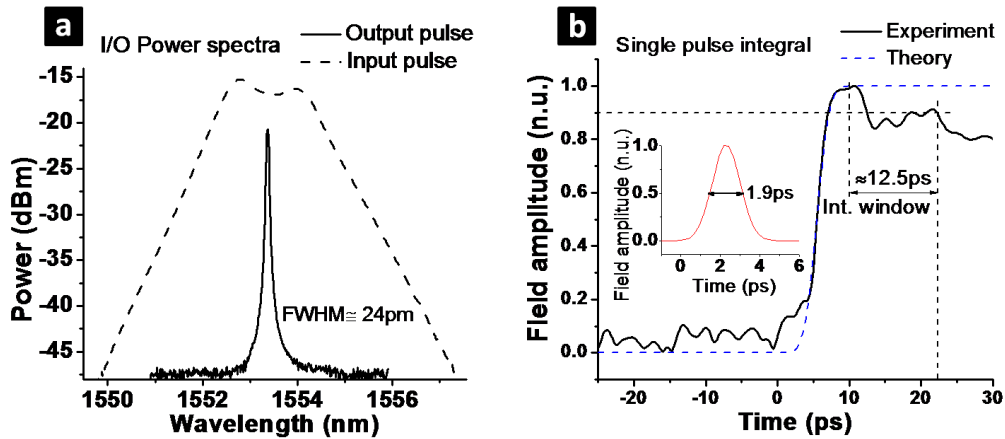

Fig. 4. (a) Optical spectra of the laser output (input to the integrator, dashed line) and output of the resonator (solid line) recorded at the drop port with the operative ring resonance excited at the input port by the laser. (b) the experimentally measured temporal profile of the output of the device (black solid curve), representing the 1st-order integral of the laser pulses, as well as the theoretically calculated integral (blue dashed curve). The inset in (b) (red curve) is the input laser pulse temporal profile. All experimental measurements were obtained using the FTSI based approach described in the text.

Figure 5 shows the results of the second set of experiments, where the Michelson interferometer was used to generate two more input signals, represented by the red curves in the insets of Fig. 5-a,c. These two pulse-train waveforms consisted of in-phase (Inset Fig. 5-a) and π-phase-shifted (Inset Fig. 5-c) pulse couplets, respectively. By coarsely varying the optical path difference of the interferometer, the temporal distance between these pulses was set to ~30ps for both the in-phase and π-phase-shifted pulses. The phase shift between pulses was precisely fine-tuned to produce either zero or π-phase-shifts by a piezo-controller mounted on one of the two mirrors of the pulse shaper. The magnitude of the phase shift was monitored by measuring the optical spectral interference between the two pulses (for example, a π (zero)-phase shift between the two delayed pulses translates into a zero (peak) in the spectral interference pattern at the pulses' central frequency).

The main plots in Fig. 5 represent: (a) the 1st- and (b) 2nd-order integrals of the in-phase signal and (c) the 1st- and (d) the 2nd-order integrals for the π-shifted pulse train. For all plots, the solid black curves are the experimentally measured temporal profiles, as directly recovered from the experimental spectra. The dashed blue lines correspond to the theoretical cumulative time integrals of the corresponding ideal input waveforms. Finally, the solid green lines are temporal convolution products. For the case of the 1st order integral, the convolution was performed between the corresponding ideal input and the impulse response of Fig. 4-b. For the 2nd order integral, the convolution is performed between the FTSI-recovered signal of the 1st integral and the same impulse response. The convolution plots, besides being a further demonstration of the robustness of our system, also prove our original assumption that the 1st-order integral (black solid curve) reported in Fig. 4-b can be considered the impulse response of our system. In fact, only under this condition we can have good agreement between the convolution plots and the experimentally calculated integrals. All the input waveforms represented by the red curves of the insets in Fig. 4-b and Fig. 5-a,c are ideal Gaussian approximations of the experimentally measured input signals, and they were used to evaluate the correspondent theoretical cumulative time integral.

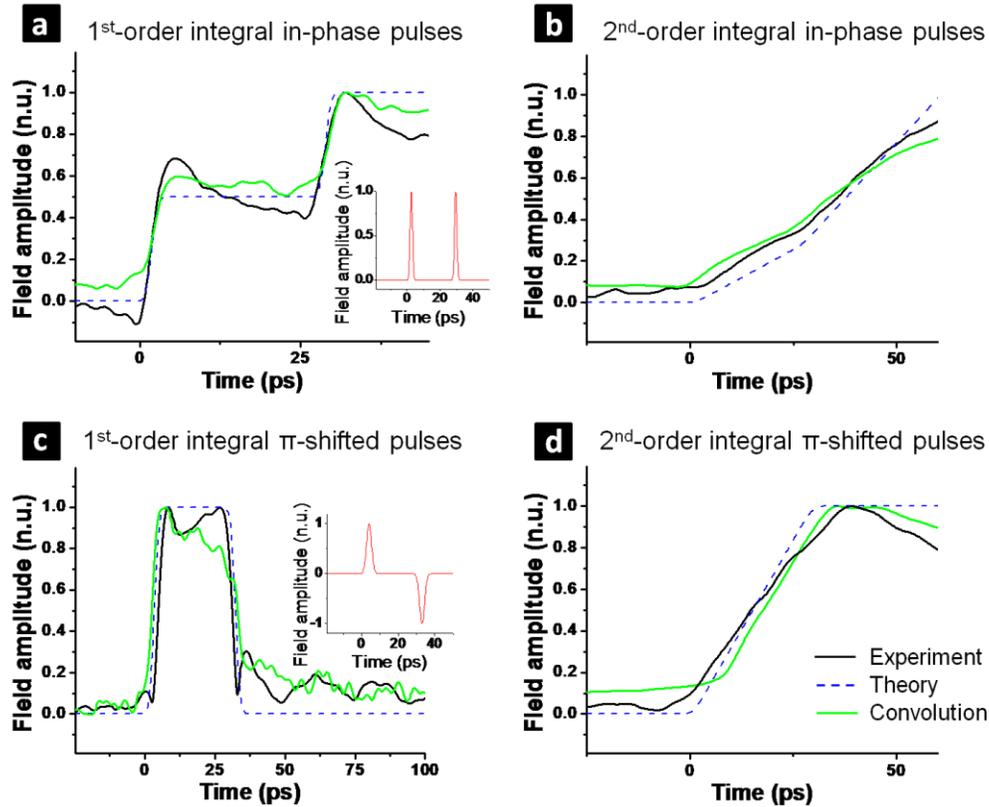

Fig. 5. Experimental results. (a) the 1st- and (b) 2nd-order integrals of the in-phase signal; (c) the 1st- and (d) the 2nd-order integrals for the π-shifted pulse train. For all these plots, the solid black curves represent the experimentally measured temporal profiles obtained using the FTSI based method described in the text. The dashed blue lines correspond to the theoretical cumulative time integrals. Finally, the solid green lines are temporal convolution products. For the case of the 1st order integral, the convolution was performed between the corresponding ideal input and the impulse response of Fig. 4-b. For the 2nd order integral, the convolution is performed between the FTSI-recovered signal of the 1st integral and the same impulse response. All the input waveforms represented by the red curves of the insets in Fig. 4-b and Fig. 5-a,c are ideal Gaussian approximations of the experimentally measured input signals, and they were used to evaluate the correspondent theoretical cumulative time integral.

These results clearly show that this device effectively acts as a complex-field cumulative temporal integrator with an operational bandwidth that exceeds 400GHz. The signal profiles were chosen to demonstrate the ability to perform important functions such as ultra-fast optical counting (in-phase pulses) and memories (π-shifted pulses) [7, 9]. The 2$^{nd}$ order integrals also fit quite well to both the theoretical cumulative integral and the convolution signal, thus proving the robustness of this device. One of the most efficient ways to perform higher order optical integration is by using FBG-based components [6]. Our approach largely avoids the need to concatenate identical fundamental (1$^{st}$-order integration) devices to achieve higher order integration, as well as drastically reducing the technological requirements that would otherwise be needed with FBG based approaches. Our results demonstrate that the approach of using micro ring cavities to attain higher order integration is an effective scheme to maximize scalability while at the same time minimizing the complexity of the device.

While the relatively limited integration window of our low-Q resonators may still preclude their use for some applications such as optical memories [9], their integration bandwidth compares very favorably with state-of-the-art electronic systems. When combined with their simple design, well accepted fabrication processes, and range of other functions that can be realized in this platform [24-27], this demonstrates their potential as fundamental components for future ultra-fast optical processing circuits.

## 4. Conclusions

We report a 1$^{st}$- and 2$^{nd}$-order on-chip ultra-fast all-optical integrator, achieving a processing speed greater than 400GHz (corresponding to a time resolution of ~1.9ps). By adopting comparatively low-Q cavities (Q~65.000) we improve the device throughput as well as the processing speed, thus demonstrating that higher order integration can be obtained with almost no increase in the overall device complexity. This work represents an important step towards the realization of efficient optical signal-processing circuits capable of overcoming the limitations in bandwidth and power consumption imposed by electronics.

### Acknowledgments

This work was supported by the Natural Sciences and Engineering Research Council of Canada (NSERC) - contract no. PDF-387780-2010 and the Australian Research Council (ARC) Discovery Projects and Centres of Excellence programs.